\documentclass[preprint,aps,prl,psfig]{revtex4}
\usepackage{amsmath}
\usepackage{graphicx}
\newcommand{\be}{\begin{eqnarray}}
\newcommand{\ee}{\end{eqnarray}}

\begin{document}
 
\title{Exotic dense matter states pumped by relativistic laser plasma in the radiation dominant regime}

\author{J. Colgan$^1$, J. Abdallah, Jr.$^1$, A. Ya. Faenov$^{2,10}$, S. A. Pikuz$^2$, E. Wagenaars$^3$, N. Booth$^4$, C. R. D. Brown$^5$,
O. Culfa$^3$, R. J. Dance$^3$, R. G. Evans$^6$, R. J. Gray$^7$, D. J. Hoarty$^5$, T. Kaempfer$^8$, K. L. Lancaster$^4$, P. McKenna$^7$,
A. L. Rossall$^3$, I. Yu. Skobelev$^2$, K. S. Schulze$^8$, I. Uschmann$^8$, A. G. Zhidkov$^9$, and N. C. Woolsey$^3$}
\affiliation{$^1$Theoretical Division, Los Alamos National Laboratory,
         Los Alamos, NM 87545, USA \\
$^2$Joint Institute for High Temperatures, Russian Academy of Sciences, Moscow 125412, Russia \\
$^3$York Plasma Institute, Department of Physics, University of York, York YO10 5DD, United Kingdom \\
$^4$Central Laser Facility, STFC Rutherford Appleton Laboratory, Didcot, Oxfordshire OX11 0QX, United Kingdom \\
$^5$AWE, Aldermaston, Reading, Berkshire RG7 4PR, United Kingdom \\
$^6$Department of Physics, Imperial College, London SW7 2AZ, United Kingdom \\
$^7$SUPA, Department of Physics, University of Strathclyde, Glasgow G4 ONG, United Kingdom \\
$^8$Institut f\"ur Optik und Quantenelektronic, Friedrich-Schiller-Universit\"at Jena, Max-Wien Platz 1, D-07743 Jena, Germany \\
$^9$PPC Osaka University and JST, CREST, 2-1, Yamadaoka, Suita, Osaka 565-0871, Japan \\
$^{10}$Quantum Beam Science Directorate, Japan Atomic Energy Agency, Kizugawa, Kyoto 619-0215, Japan
}
 
\begin{abstract}

{\bf 

The properties of high energy density plasma are under increasing scrutiny in recent years due to their importance
to our understanding of stellar interiors, the cores of giant planets$^{1}$, and the properties of hot plasma
in inertial confinement fusion devices$^2$. 
When matter is heated by X-rays, electrons in the inner shells are ionized before the valence electrons.
Ionization from the inside out creates atoms or ions
with empty internal electron shells, which are known as hollow atoms (or ions)$^{3,4,5}$.
Recent advances in free-electron laser (FEL) technology$^{6,7,8,9}$ have made possible the creation of condensed matter consisting
predominantly of hollow atoms.
In this Letter, we demonstrate that such exotic states of matter, which are very far from equilibrium, can also be formed by more 
conventional optical laser technology when the laser intensity approaches the radiation dominant regime$^{10}$. 
Such photon-dominated systems are relevant to studies of photoionized plasmas found in active galactic nuclei and X-ray binaries$^{11}$.
Our results promote laser-produced plasma as a unique ultra-bright x-ray source for future studies of matter in extreme
conditions as well as for radiography of biological systems and for material science studies$^{12,13,14,15}$.
}
 
\end{abstract}

\maketitle

 
\bigskip

A high-intensity laser rapidly ionises valence electrons from a thin target through field ionisation$^{16}$. 
These electrons are accelerated in the strong electric field of the laser and can quickly reach very high, relativistic, energies.
During the interaction of an intense laser pulse with a thin foil some of these hot electrons 
oscillate through the foil$^{16,17}$, due to reflection by the fields on each surface, known as refluxing. 
These electrons radiate X-rays via Thomson scattering of the
incident and scattered laser light and via Bremsstrahlung through the strong plasma field at the foil 
surfaces (see Fig.~1). In the case of non-linear Thomson scattering an intensity and spectrum 
estimate is possible$^{10,18,19}$ and for a laser intensity of approximately $5\times 10^{20}$ W/cm$^2$, this yields an
X-ray intensity of around $10^{19}$ W/cm$^2$ and in the keV range.  
This is well beyond the perturbative regime for laser-matter interactions.
Estimates of the Bremsstrahlung induced by the plasma electric field$^{20}$ 
results in an X-ray radiation intensity of $\sim$ $3\times 10^{19}$ W/cm$^2$ for similar 
laser intensity and a laser spot size of 10~$\mu$m. We note that modern X-ray
FELs produce extremely bright monochromatic and coherent X-ray fluxes, in comparison 
this radiation field is polychromatic and up to two orders of magnitude more intense. 

We therefore find that long-wavelength, super-intense laser radiation (of order $> 10^{20}$ W/cm$^2$) drives the ionized electrons
from the target to very high energies, which thermalize via intense emission of X-ray photons. The resulting X-ray radiation field 
is so intense that hollow atoms dominate, resulting in the creation of exotic states of matter, as illustrated in the lower panel of Fig.~1. 
Such radiation dominated exotic states are examples of atomic systems far from equilibrium. 
This is the first observation of such states driven by polychromatic radiation fields with intensities of order $10^{19}$ W/cm$^2$,
which replicate the processes that may occur in active galactic nuclei (AGN) where intense X-ray fields drives inner shell photoexcitation$^{11}$.
This process is demonstrated in this joint theoretical and experimental study.

The measurements were made at the Vulcan Petawatt Laser Facility at Rutherford Appleton Laboratory. 
Spectrally resolved X-ray emission between 7 and 8.4 \AA~ from pure aluminium foils of
1.5 and 20 $\mu$m thicknesses was investigated. 
Al $K$-shell spectra measured at different laser and target parameters are shown in the lower panel of Fig.~2. 
The accepted wavelengths for the Ly$_{\alpha}$, He$_{\alpha}$, and $K_{\alpha}$ lines are indicated by the vertical dashed lines. 
The measured resonance line centers are shifted to longer wavelength with respect to the accepted values by 10 and 20-25  m\AA~ respectively.
This is due to ion acceleration and subsequent Doppler shifts$^{21}$.
The black curve in the lower panel represents the data obtained from a 1.5 $\mu$m thin Al foil irradiated with 160~J 
on target, the maximum available. For this case the laser intensity at the central area of the focal spot of 8 $\mu$m diameter reached the value of 
$2\times10^{20}$ W/cm$^2$.
Intense hollow atom spectral lines are observed in the spectral range between the Ly$_{\alpha}$ and He$_{\alpha}$  resonance lines, i.e.
between 7.17 and 7.76 \AA, and also on the short wavelength side (blue edge) of the K$_{\alpha}$ line just below 8.34 \AA. 

As shown in Fig.~2 the reduction of the laser pulse energy from 160~J to 64~J 
(and corresponding decrease in laser intensity at the target surface) 
leads to remarkable changes in the spectra - a substantial decrease in the yield of hollow ion spectral lines occurs. 
For a measurement made using
the full laser energy but a 20~$\mu$m thick target (blue curve), we observe a similar change in the spectra.
For reference, the data obtained$^{22}$ using a much lower laser energy of 80 mJ (laser intensity of $5\times10^{17}$ W/cm$^2$) 
is also presented in Fig.~2 (lowest curve). Such a spectrum is typical for a low-Z target heated by a moderate intensity high contrast
laser pulse. No hollow ion transitions are observed in such measurements.

It is clear that the large energy deposited in the present experiment leads to a quite different physical picture than that used 
for lower energy laser experiments.
The intense Vulcan laser rapidly ionizes valence electrons from the Al target through field ionization$^{16}$. 
These electrons are accelerated in the strong electric and magnetic fields of the laser, and
the ponderomotive electron energy$^{23}$ 
in a laser field at an intensity
of $10^{20}$ W/cm$^2$, is around 10~MeV. The thin Al target is essentially transparent to 
these highly-relativistic electrons$^{16}$, but the electrons quickly 
lose energy through Bremsstrahlung$^{24}$ (and (non-linear) Thomson
scattering$^{18,19}$). Bremsstrahlung emission$^{24}$ at such relativistic electron energies results in a reasonably flat
distribution of photons in the X-ray energy regime.
These photons interact with the Al ions and atoms in the target by interacting with electrons in the inner atomic shells and 
ejecting multiple $K$
and $L$-shell electrons. Inner-shell electrons are preferentially ionized by high-energy photons so that
the excess recoil momentum can be absorbed by the nucleus. This is demonstrated in Fig.~3, where we show the ratio of $K$ to $L$ shell
ionization probability as a function of the photon (pink line) and electron (blue line) energy. 
It is clear that high-energy photons at 1~keV or higher are much
more efficient at removing $K$ shell electrons than electrons of similar energies. 
A previous measurement$^{25}$, which used similar laser intensities to the present
measurement but a much thicker target,  found that
electron pumping of hollow ions was the main ionizing mechanism. In this case it was found that the generation of hollow ions was not efficient.

After photoionization, the $K$-shell is rapidly filled by
Auger decay, resulting in the further emission of $L$-shell electrons. Radiative decay (where an $L$-shell electron drops back
into the $K$-shell) also accounts for a small fraction of the recombination, and results in observable emission lines usually referred to as
satellites.
We observe different emission spectra depending on the laser energy and spot size. For the lower laser energy case (green
line in figure~2 at 64~J), we observe little emission, with prominent peaks just below $7.8$ and $7.2$ \AA. These lines
correspond to the He$_{\alpha}$ and Ly$_{\alpha}$ transitions. 
Some satellites to these lines can also be observed.
In the full laser
energy case (160~J; black line in Fig.~2) the spectra is dramatically different. 
The `usual' He$_{\alpha}$ and Ly$_{\alpha}$ lines are dominated
by significant emission between $7.3$ and $7.7$ \AA~ and between $7.9$ and $8.3$ \AA. The new emission at
wavelengths lower than the He$_{\alpha}$ line is due to the radiative
decay from the $L$-shell to the $K$-shell within ions which have a double $K$-shell vacancy, as well as several $L$-shell
vacancies. A possible hollow atom configuration is shown schematically in the inset of Fig.~3.
Emission arises from many ion stages of Al, from Al III through Al X. For emission to be observed from such ion stages
at these high field intensities implies that the electron density remains high, so that three-body recombination drives the
ions back towards the neutral stage. This indicates that the radiation field is 
interacting with near-solid density Al, that is at electron densities greater than $10^{23}$ cm$^{-3}$ and ion densities
$>10^{22}$ cm$^{-3}$.
 The emission in the longer wavelength range
($7.9$ to $8.3$ \AA) results from a similar radiative decay process, except that in this case 
the initial state has only one $K$-shell vacancy. This emission occurs at slightly later times, when the electron density
is smaller and the radiation field has significantly decreased.
Such transitions are less energetic and so occur at longer wavelengths. Refluxing of fast electrons, which are
responsible for x-ray radiation, between the target's front and rear occurs only in thin foils and vanishes in thick ones.
This is the primary reason for the different spectra observed from the 1.5 $\mu$m and 20$\mu$m foils.

Our measurements are analyzed by performing sophisticated atomic kinetics calculations using the ATOMIC code$^{26}$. 
The upper panel of Fig.~2 shows calculations (all using ATOMIC), modeled by a
plasma exposed to a radiation field peaked at 3 keV, with a bulk electron temperature of 55 eV
and electron density of $3\times 10^{23}$ cm$^{-3}$. A small fraction (5\%) of the electron distribution was in a hot-electron tail with
a temperature of 5 keV. 
The full calculation is the lower pink line and we find clear evidence of emission from hollow ions from a range of Al charge states.
The comparison with the measured spectrum (black/pink lines at the top of the lower panel of Fig.~2) shows that the kinetics calculation successfully
reproduces almost all features of the measurement.
We find that transitions involving double $K$-shell holes (between 7.2 and 7.7 \AA)  are visible for a range of
Al ions from Al IV (neon-like) through Al IX (boron-like). Interestingly, the peak heights corresponding to the transitions associated with each ion stage
are in fair agreement with the broad charge-state distribution of the ions at these conditions shown in the Fig.~2 inset.
Emission lines have been identified that arise from ions with up to six electrons removed from the $K$ and $L$ shell - true hollow ions.
We also observe significant emission at higher wavelengths (7.7 and 8.3 \AA) from single $K$-shell holes, again from a range of Al ions as indicated.
This emission is mostly from plasma at lower densities interacting with a much reduced radiation field.

To unambiguously demonstrate that the intense radiation field is responsible for the double $K$-shell hollow ion emission, we also show
two calculated spectra (grey and red lines) in the upper panel of Fig.~2, with no radiation field included in the calculation. 
Virtually no emission lines due to 
hollow ions are observed, particularly in the region between 7.2 and 7.7 \AA.
The red line shows the influence of the 5~keV hot-electron tail on the emission spectrum.
It is clear that the radiation field has a dramatic effect.
In Fig.~2 the blue line shows the effect of radiation on the calculation whilst omitting the hot-electron tail. 
We find that the hot electrons do excite hollow ions, 
but provide only a small part of the measured intensity of the hollow ion spectral lines; Fig.~3 provides an explanation for this.

Our study demonstrates that with high-powered laser technology 
signatures of the radiation dominant regime are apparent and polychromatic X-ray fields have  
extreme intensity which compliment the monochromatic and coherent fields delivered by XFELs. 
We show that polychromatic X-ray fields drive states of matter far from equilibrium reproducing some of the conditions that may occur in AGNs.
In this case the
radiation dominant regime of matter ionization is demonstrated, that in turn allows us to observe radiation 
involving double $K$-shell hole states, true hollow atoms. 
We plan further studies of the properties of these exotic states of matter created in super-intense laser pulse experiments.
Finally, our results demonstrate how laser-plasma interaction physics will change as laser intensities move
towards the radiation dominant regime, a region where classical
electrodynamics begins to break down$^{27,28,29}$.

\section{Methods Summary}

The Vulcan Petawatt Laser Facility at Rutherford Appleton Laboratory provides a beam using optical parametric 
chirped pulse amplification (OPCPA) technology with a central wavelength of 1054~nm and a pulse 
full-width-half-maximum (FWHM) duration of 0.7~ps. The OPCPA approach enables an amplified spontaneous 
emission (ASE) to peak-intensity contrast ratio exceeding $1:10^9$
several nanoseconds before the peak of the laser pulse.  
The laser pulse  was focused with an $f/3$ off-axis parabola 
providing up to 160~J on the target. The maximum laser intensity of $3\times10^{20}$ W/cm$^2$
was achieved with a laser focus containing approximately 30\% of the energy in an 8~$\mu$m
(FWHM) diameter spot. The p-polarized laser beam was incident on target 
at $40^{\circ}$ from the target surface normal

The spectra were measured using a FSSR (focusing spectrometer with spatial resolution) equipped with spherically bent 
(radius of curvature $R=150$~mm) mica crystals. The FSSR spectrometers
were aligned to observe radiation in the wavelength range from $7.0$ to $8.4$ \AA~(energy from $1.47$ to $1.77$~keV) 
containing the K-shell spectra from multi-charged and neutral Al ions. The spectra were acquired from the
front, laser irradiated surface, at 45$^{\circ}$ to the target normal as schematically shown in Fig.~1(c). 
Background fogging and crystal fluorescence due to intense 
fast electrons were reduced using a pair of 0.5~T permanent magnets that formed a slit 10~mm wide in front of the crystal. 
Spectra were recorded on Kodak Industrex AA400 photographic X-ray film.

Preliminary time-dependent
calculations using a simplified atomic model were performed to determine the bulk plasma conditions. We then performed detailed steady-state
calculations at various plasma conditions to compute the emission spectrum as a function of electron temperature and density and radiation
temperature, where we assumed that the radiation field seen by the plasma was a Planckian distribution. Our preliminary time-dependent
calculations show that, at the high electron densities
under consideration here the system reaches steady-state very quickly.
Our calculations include
all relevant atomic processes, that is, photoionization, collisional ionization, autoionization, collisional and radiative excitation and de-excitation,
and all recombination processes. The atomic structure and collisional cross section calculations were made using the Los Alamos suite of 
atomic physics codes$^{30}$. The resulting rate
equations are solved to produce a set of ion populations that depends sensitively on the electron temperature and density, the radiation
field, and the fraction of energetic electrons included in our model. Opacity effects are included via escape factors. The resulting
level populations are used to produce emission spectra, where the effects of detailed lines are included via a mixed-UTA (unresolved-transition-array)
approach. To correctly model the exotic plasma created experimentally, it is necessary to include a very large number of configurations,
representing Al ions in which up to six electrons have been removed from the $K$ and $L$ shells and placed in excited states. Such multiply excited
states were found to be important in accurately reproducing the observed emission spectra. In particular, we find that the radiative decay
from the $L$ to $K$ shell in any one specific configuration results in a relatively weak line, 
but that the very large number of configurations with different combinations
of spectator electrons in excited states, in which this radiative transition occurs, allows the observed transition to become prominent.

\section{Acknowledgments}

We thank the Vulcan technical and target preparation teams at the Central Laser Facility for their support during the experiments. 
The research leading to these results has received funding from the Science and Technology Facilities Council, 
and the Engineering and Physical Science Research Council (Grant No. EP/E048668/1) of the United Kingdom.
The Los Alamos National Laboratory is operated by Los Alamos
National Security, LLC for the NNSA of the U.S. DOE under Contract No.
DE-AC5206NA25396.
The work of the JIHT RAS team was supported by RF President Grant \# MK-4725.2012.8 and RAS Presidium Program for Basic Research \#2.

\section{Author Contributions}

N.C.W., E.W., N.B., R.G.E., I.U and S.A.P. conceived this project. Interpretation of the spectroscopic data was carried out 
by J.C., J.A, A.Y.F., and I.Y.S. The experimental data was analysed by A.Y.F., S.A.P., and R.J.D. 
The Vulcan experiment was carried out by E.W., O.C., R.J.D., A.K.R., C.R.D.B., S.A.P, K.S.S, T.K., I.U., N.B, K.L.L., R.J.G., and N.C.W. 
Additional theoretical support was provided by A.G.Z. and additional experimental support was provided by P.M. and D.J.H. 
All authors contributed to the work presented here and to the final paper.

\newpage
\begin{figure}
\label{fig1}
\includegraphics[scale=1]{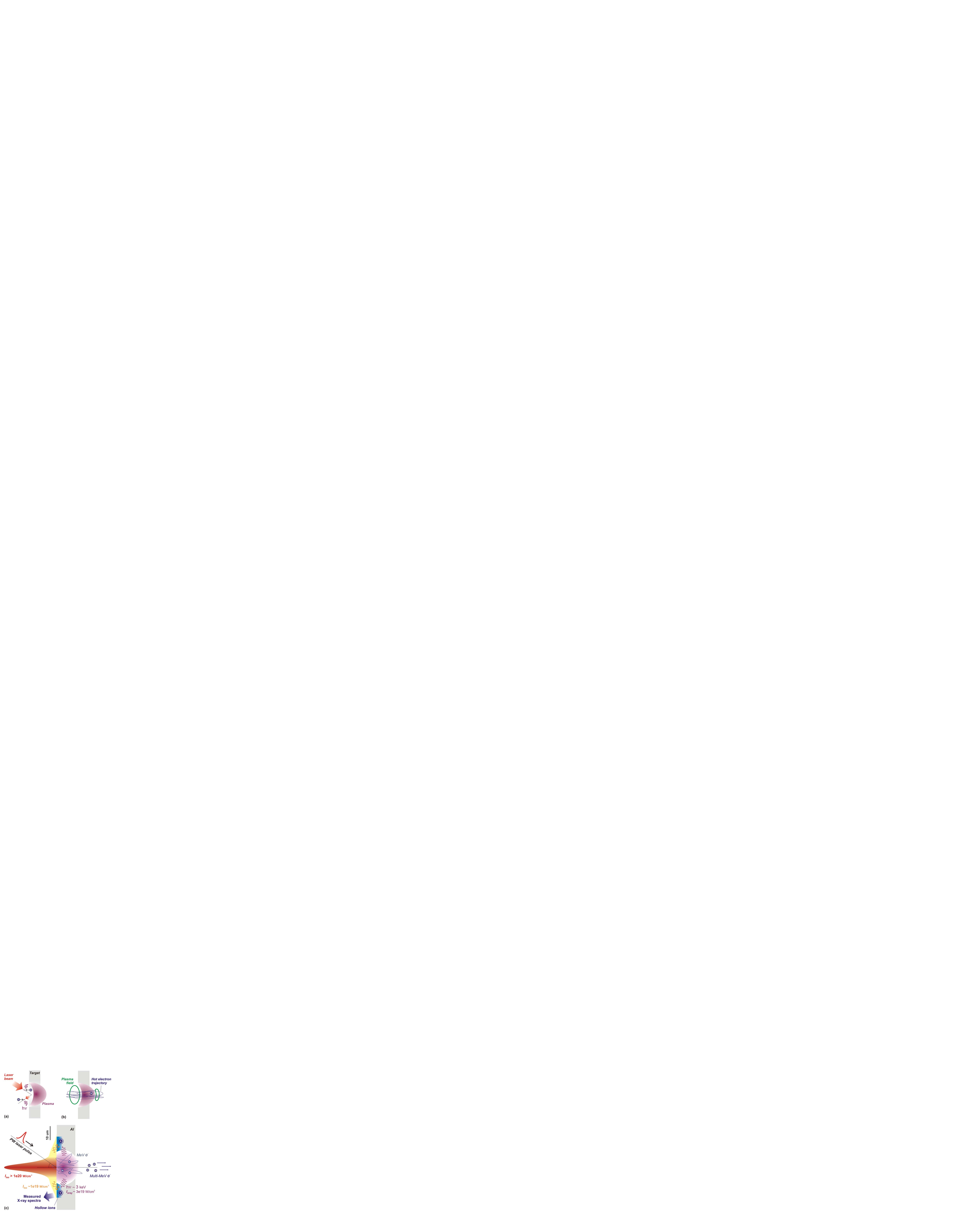}
\caption{Bright X-ray radiation may be produced during the interaction of the intense laser pulse
with the target via Thomson scattering, as illustrated in panel (a), or through
Bremsstrahlung on the surface, panel (b). The lower panel (c) shows a schematic diagram of the formation
of hollow atoms by the intense laser.
}
\end{figure}

\newpage
\begin{figure}
\label{fig2}
\includegraphics[scale=0.8]{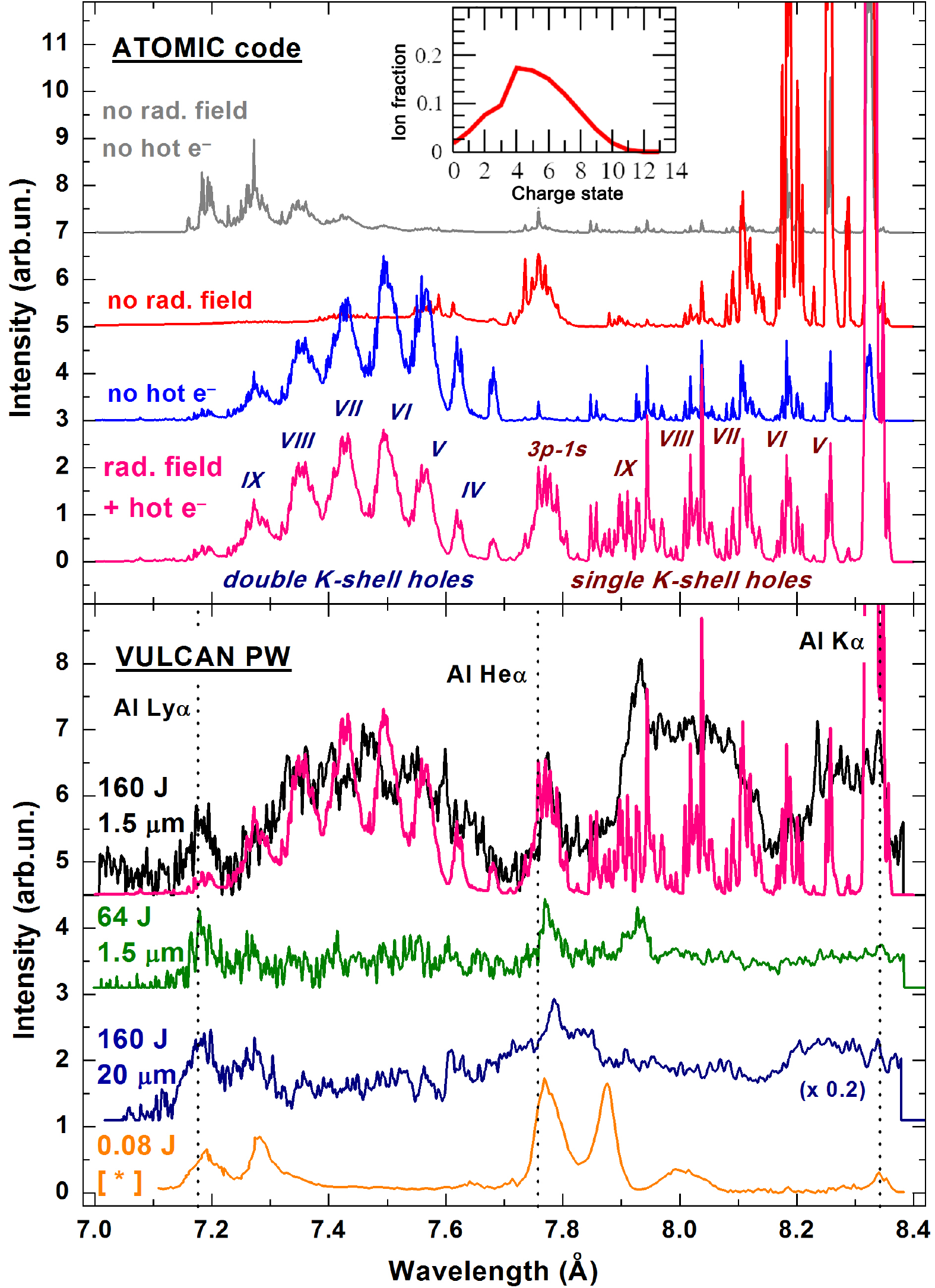}
\caption{ 
Aluminum K-shell spectra calculated by ATOMIC code (upper panel) and measured at Vulcan PW laser (lower panel).
ATOMIC calculations use an electron temperature of 55 eV and electron density of $3\times 10^{23}$ cm$^{-3}$
with different options as indicated.
The inset shows the ion charge distribution for the full calculation (pink line in both panels) that
includes the radiation field and hot electrons.
Experimental spectra were measured for Al foil targets of 1.5 $\mu$m or 20 $\mu$m thickness and with 
laser energies of 160 or 64~J. 
As a reference the lowest curve represents the data obtained using a 150 fs 80 mJ laser pulse at $5\times 10^{17}$ W/cm$^2$ intensity$^{25}$.
The vertical dashed lines indicate the Al Ly$_{\alpha}$
He$_{\alpha}$ resonance transitions and the Al K$_{\alpha}$ transitions.
All plots are offset vertically, and the data for the 20 $\mu$m Al target 
is multiplied by a factor of 0.2, for better visibility and convenient comparison.  We compare the spectrum obtained at 160~J and a thickness 
of 1.5 $\mu$m with the full ATOMIC calculation. 
}
\end{figure}

\newpage
\begin{figure}
\label{fig3}
\includegraphics[scale=1]{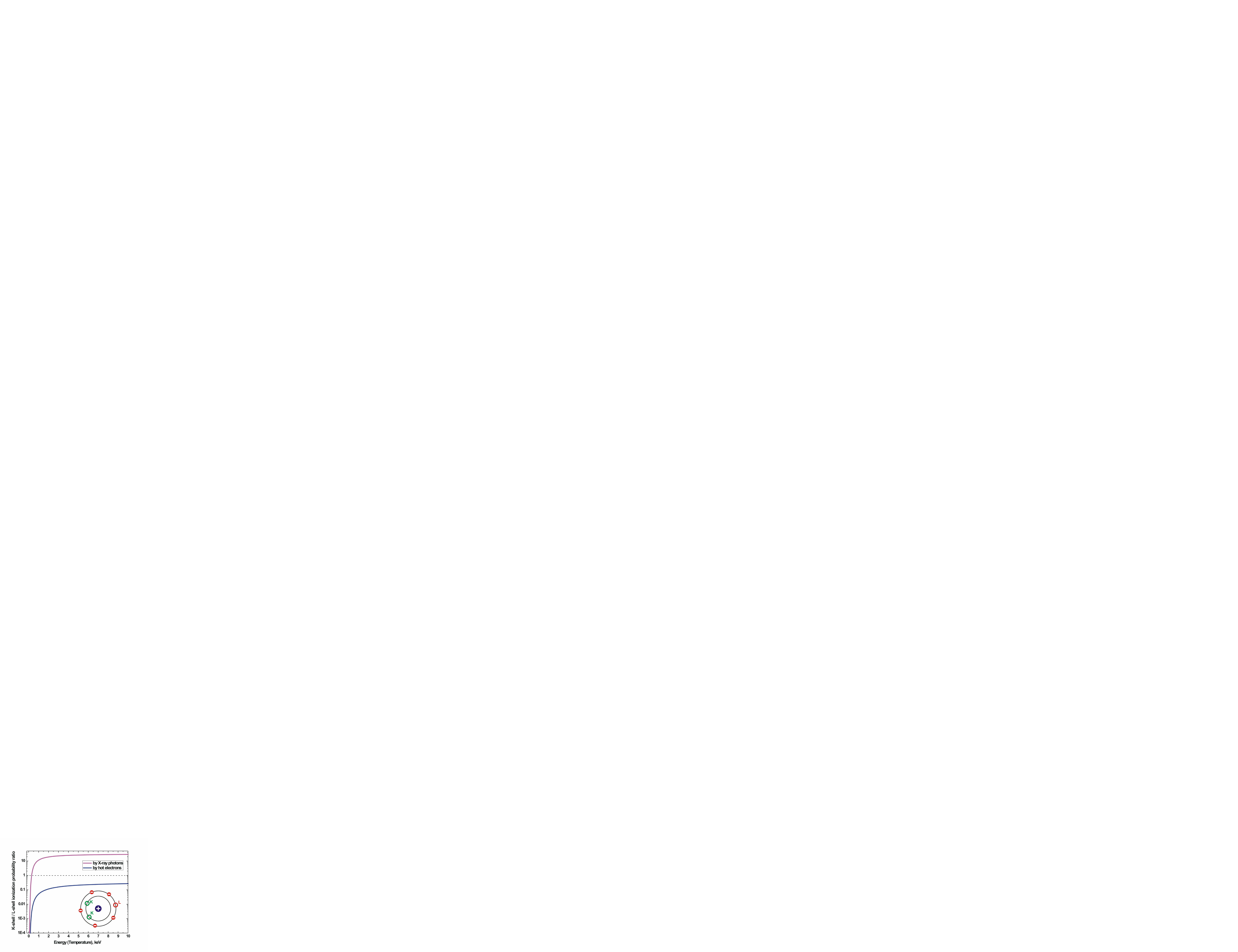}
\caption{Ratios of probabilities of ionization from the $K$ and $L$ shells. Pink line: photo-induced
ionization; blue line: electron-induced ionization. Inset: schematic representation of a hollow ion with
several $K$ and $L$-shell vacancies and populated outer shells.
}
\end{figure}

\end{document}